\documentclass[12pt,a4paper]{amsart}
\usepackage{amsmath,amsthm,amstext,amssymb,a4}
\usepackage[T1]{fontenc}

\usepackage{graphicx}

\theoremstyle{plain}

\renewcommand{\.}{{\cdot}}

\newtheorem{Def}{Definition}[section]

\newtheorem{Thm}{Theorem}[section]

\newtheorem{corollary}{Corollary}[section]

\newtheorem{Prop}{Proposition}[section]

\begin{document}
\author[]{A. Dirmeier \and M. Plaue \and M. Scherfner}
\address{A. Dirmeier, M. Plaue and M. Scherfner: Institute of Mathematics, Technische Universit{\"a}t Berlin,
Str.~d.~17.~Juni 136, 10623 Berlin, Germany}

\title[]{On Conformal Vector Fields Parallel to The Observer Field}

\begin{abstract}
\noindent  We review a theorem by Hasse and Perlick establishing a result characterizing parallax-free cosmological models via three equivalent properties -- namely the existence of a redshift potential, the existence of a conformal vector field parallel to the observer field, and the vanishing of the shear of the observer field together with some integrability condition. We are able to provide a simplified proof using Noether's theorem to calculate a conserved quantity along lightlike geodesics that is connected with the conformal symmetry. Thereby we derive more detailed information about the connection of the kinematical invariants to the redshift isotropy and the connection of conformal vector fields to the causality of spacetime. 
\end{abstract}

\maketitle


\section{Introduction}
In 1985, Treder \cite{Treder1985} proposed that cosmic rotation could be observed as an aberrational motion of distant cosmic objects. Following this idea, Hasse and Perlick proved 1989 (see \cite{Hasse1988}) that a non-vanishing cosmic rotation cannot be the only cause of parallax effects. According to their analysis, the only kinematical property of the model that is directly related to parallax effects is the shear. However, we will show that the rotation cannot be totally arbitrary. This theorem (called Hasse--Perlick theorem in this article) also shows that a cosmological model ist parallax-free if and only if there exists of a conformal vector field proportional to the observer field which is in turn equivalent to the existence of a redshift potential.

We will use the term ``parallax'' in its most general sense as the general relativistic amalgamation of ``proper motion'', ``aberration'' and ``parallax''.

In this article, after some mathematical preliminaries, we prove some fundamental properties of conformal vector fields. A simple but important observation turns out to be the fact that constant multiples of a conformal vector field constitute a natural equivalence class of conformal vector fields. This will lead to an important proposition which states that a family of conformal vector fields proportional to the observer field---each defined on an open subset of a manifold---leads to an equivalence class of globally defined conformal vector fields, if the manifold can be covered by these open subsets.

Furthermore, all physical quantities only depend on the equivalence class of conformal vector fields and not on a chosen representative. We will illustrate this fact in the case of the redshift and the angle between light rays.

In the third section, we give a simplified proof of the Hasse--Perlick theorem and discuss the kinematical constraints of a parallax-free spacetime.

Finally, we obtain some results connecting the causality of a spacetime with the existence of a conformal vector field with certain kinematical properties.

\section{Preliminaries}\label{sec:prel}
\subsection*{Notation and Conventions}
In this article, we will take $M$ to be a four-dimensional paracompact $C^{\infty}$-manifold. Also, we will assume any function or tensor field on $M$ to be smooth.
The tangent and cotangent bundle over a manifold $M$ will be denoted by $TM$ and $T^{*}M$ respectively.
For a diffeomorphism $\varphi\colon M\to M$, we denote the pullback map by $\varphi_{*}$ and the push-forward map by $\varphi^{*}$.

The metric tensor will be written as a $C^{\infty}$-map $g\colon TM\times TM\to\mathbb{R}$. Vector fields will be denoted by capital letters ($X,Y,\ldots$). An index on such a vector field denotes the vector field evaluated at a point $p\in M$ (e.g.\ $X_{p}\in T_{p}M$). Conformal vector fields will usually be denoted by greek letters $(\xi,\eta,...)$. The corresponding covector to the vector $X\in TM$ will be denoted by $g(X,\cdot)\in T^{*}M$. For general $k$-forms, we will use latin letters ($a,b,\ldots$). Although we prefer invariant notation, a (pseudo-) orthonormal basis will be denoted by $\{E_{i}\}=\{E_{0},E_{\alpha}\}$ with latin indices $i,j,\ldots=0,1,2,3$ and greek indices $\alpha,\beta,\ldots=1,2,3$, such that $g(E_{\alpha},E_{\beta})=\delta_{\alpha\beta}$, $g(E_{0},E_{0})=-1$ and $g(E_{0},E_{\alpha})=0$. Components with respect to a local chart we will denote with latin indices $a,b,\ldots=0,1,2,3$, therefore a point $p\in M$ has components $p^{a}$. Similarly we denote vector and covector components with respect to a natural chart by $X=X^{a}\frac{\partial}{\partial x^{a}}$ and $g(X,\.)=X_{a}dx^{a}$ respectively.

By $\rfloor$ we will denote the inner product of a vector with a $k$-form, i.e.\ $(X\rfloor a)(\.,\ldots,\.)=a(X,\.,\ldots,\.)$.

By $\otimes$ we will denote the ordinary tensor product, by $\vee$ the symmetric tensor product and by $\wedge$ the antisymmetric tensor product (wedge product). For two vectors $X$ and $Y$ we define these by

\begin{eqnarray*}
	X\vee Y & = & X\otimes Y+Y\otimes X\\
	X\wedge Y & = & X\otimes Y-Y\otimes X.
\end{eqnarray*}

Similarly, by $\textrm{sym}(\cdot)$ we denote the symmetric part of a tensor of type $(r,0)$ or $(0,s)$ and by $\textrm{antisym}(\cdot)$ its antisymmetric part. For example, for a tensor $T$ of type $(0,s)$, which has components $T_{a_{1}\dots a_{s}}$, these are defined by

\begin{eqnarray*}
	\textrm{sym}(T_{a_{1}\dots a_{s}}) & = & \frac{1}{s!}\sum_{\sigma\in P(a_{1}\dots a_{s})}T_{\sigma(a_{1}\dots a_{s})}\\
	\textrm{antisym}(T_{a_{1}\dots a_{s}}) & = & 
	\frac{1}{s!}\sum_{\sigma\in P(a_{1}\dots a_{s})}\textrm{sgn}(\sigma)T_{\sigma(a_{1}\dots a_{s})},
\end{eqnarray*}
where $P(a_{1}\dots a_{s})$ is the group of permutations of the indices $a_{1}\dots a_{s}$, and $\textrm{sgn}(\sigma)$ denotes the sign of the permutation $\sigma$.

We denote by $\nabla_{X}$ the covariant derivative along a vector field $X$ with respect to the Levi--Civita connection on the manifold $M$ which obeys the usual derivation rules. If $X$ is the tangential vector field along a curve in $M$ parametrized by $s$, we also denote the covariant derivative by $\nabla_{X}=\frac{D}{ds}$.

By $L_{X}$ we denote the Lie derivative along any vector field $X$. The Lie derivative obeys the usual derivation rules among which
\begin{equation}\label{eqn:liedec}
	L_{fX}g=fL_{X}g+df\vee g(X,\cdot)
\end{equation}
for any scalar function $f$ will be of special importance. For the Lie derivative of differential forms the following fundamental equation holds
\begin{equation}\label{eqn:lieform}
 L_{X}a=X\rfloor da + d(X\rfloor a)
\end{equation}
for any vector field $X$ and any $k$-form $a$ (Cartan's magic formula). Occasionally, we will write the Lie derivative of a vector field $Y$ with respect to another vector field $X$ as the Lie bracket
\[
	[X,Y]:=L_{X}Y.
\]
The Lie algebra of all vector fields on a manifold $M$ will be denoted by $\mathcal{X}(M)$.

\begin{Def}\label{def:lorentz}
\emph{(Spacetime)} A spacetime is a tupel $(M,g)$ consisting of a four-dimensional $C^{\infty}$-manifold $M$ and a metric tensor field $g$ on $M$ with trace $tr(g)=\pm 2$, called the \emph{signature} of $g$.
\end{Def}
We will choose the signature of $g$ to be $+2$ such that a timelike vector $X$ has negative squared norm $g(X,X)<0$.

\begin{Def}\label{def:worldmodel}
\emph{(World model)} A tripel $(M,g,V)$ that consists of a spacetime $(M,g)$ and a timelike vector field $V$ (observer field) on $(M,g)$ will be called a world model.
\end{Def}
We assume the observer field $V$ to be normalized such that $g(V,V)=-1$. The existence of $V$ implies that the spacetime of a world model is time orientable. The integral curves of $V$ are called observers and may be interpreted as the world lines of the cosmic substrate (galaxies, etc.). These integral curves will usually be denoted by $\gamma_{0}$, $\gamma_{1}$, $\gamma_{2}$, etc.

\subsection*{Kinematical invariants}
It is well-known (e.g.\ \cite{Duggal1999}, chap.\ 4.1) that the covariant derivative $g(\cdot,\nabla_{\cdot}V)$ of the observer field $V$ can be decomposed into irreducible parts, called kinematical invariants
\begin{equation}\label{eqn:kininv}
	g(\cdot,\nabla_{\cdot}V)=\frac{\Theta}{3}h+\sigma+\omega-g(\nabla_{V}V,\cdot)\otimes g(V,\cdot).
\end{equation}
Here, the tensor $h$ is called \textbf{projection operator}. It projects onto the hyperplanes perpendicular to $V$
\begin{equation}\label{eqn:proj}
	h(\cdot,\cdot):=g(\cdot,\cdot)+g(V,\cdot)\otimes g(V,\cdot).
\end{equation}
It is obviously symmetric and has trace $\textrm{tr}(h)=3$.

The kinematical invariants are defined as follows:

The \textbf{volume expansion} $\Theta$
\[
	\Theta:=\textrm{div}\,V=\textrm{tr}(g(\cdot,\nabla_{\cdot}V)).
\]
The \textbf{shear tensor} $\sigma$
\[
	\sigma(\cdot,\cdot):=\textrm{sym}(g(\cdot,\nabla_{\cdot}V))+\frac{1}{2}g(\nabla_{V}V,\cdot)\vee g(V,\cdot)-
	\frac{\Theta}{3}h(\cdot,\cdot).
\]
The \textbf{rotation} or \textbf{vorticity tensor} $\omega$
\[
	\omega(\cdot,\cdot):=\textrm{antisym}(g(\cdot,\nabla_{\cdot}V))+\frac{1}{2}g(\nabla_{V}V,\cdot)\wedge g(V,\cdot).
\]
The quantity $\nabla_{V}V$ is called \textbf{acceleration} and it obeys 
\begin{equation}\label{eqn:acc}
	g(\nabla_{V}V,V)=0. 
\end{equation}
The integral curves of $V$ are geodesics if and only if the acceleration vanishes. All the kinematical invariants are quantities which live on the hyperplanes perpendicular to $V$, i.e.
\begin{equation}\label{eqn:kininvperp}
	\sigma(V,\cdot)=\omega(V,\cdot)=h(V,\cdot)=0.
\end{equation}
According to \cite{Hasse1988} one can easily check that
\begin{equation}\label{eqn:symkininv}
	L_{V}g=2\,\textrm{sym}(g(\cdot,\nabla_{\cdot}V)),
\end{equation}
using the ordinary derivation rules.

\subsection*{Conformal Vector Fields}
Now we define conformal vector fields and the conformal symmetries associated with them, as well as some of their properties.
\begin{Def}\label{def:confsym}
\emph{(1-parameter group of conformal symmetries)} On a spacetime $(M,g)$ there exists a $1$-parameter group of conformal symmetries iff there is a family of maps $\varphi_{t}:\,\mathbb{R}\times M\to M$, $(t,q)\mapsto p=\varphi_{t}(q)$ that leaves the metric conformally invariant. This means
\begin{equation}\label{eqn:confsym}
	(\varphi_{t*}g)_{q}(X_{q},Y_{q})=g_{p}(\varphi^{*}_{t}X_{p},\varphi^{*}_{t}Y_{p})=e^{\Phi t}\,g_{q}(X_{q},Y_{q})
	\qquad \forall\,X,Y\in TM,
\end{equation}
or abbreviated
\[
	\varphi_{t*}g=e^{\Phi t}\,g.
\]
$\Phi$ is a scalar function, called the conformal factor, and $t$ is the group parameter ($\varphi_{t}\circ\varphi_{s}=\varphi_{t+s}$, $\varphi_{0}=id$, $(\varphi_{t})^{-1}=\varphi_{-t}$).
\end{Def} 
Now it can be easily shown (e.g.\ \cite{Duggal1999}, chap.\ 3.4) that such conformal transformations are generated by vector fields, which are called conformal vector fields (CVF). This gives rise to the following definition.

\begin{Def}\label{def:CVF}
\emph{(Conformal vector field)} A vector field $\xi$ on $(M,g)$ is called a CVF iff there is a scalar function $\Phi$ such that the following holds:
\begin{equation}
	L_{\xi}g=\Phi g.
\end{equation}
\end{Def}

A conformal vector field $\xi$ generates a $1$-parameter group of conformal transformations since
\[
	L_{\xi}g=\lim_{t\to0}\frac{1}{t}(\varphi_{t*}g-g)=\lim_{t\to0}\frac{1}{t}(e^{\Phi t}-1)g=\Phi g.
\]
with $\varphi_{t}$ being the flow of $\xi$, which maps the manifold $M$ to itself along the integral curves of $\xi$. Because of this the angle between vectors fields along these integral curves are preserved in the following sense:

\begin{Prop}\label{prop:angle}
Let $\xi$ be a CVF on a spacetime $(M,g)$ giving rise to a $1$-parameter group of conformal symmetries $\varphi_{t}$. Let $\gamma\colon I\to M$, $t\mapsto\gamma(t)$ (with $I\subset\mathbb{R}$ being an interval, containing $0$ for simplicity, which can always be achieved by an affine reparametrization of the curve parameter), be any integral curve of $\xi$ and $X\colon I\to TM$, $t\mapsto X_{t}:=X_{\gamma(t)}$, $Y\colon I\to TM$, $t\mapsto Y_{t}:=Y_{\gamma(t)}$ two non-null vector fields along $\gamma$. At each point $\gamma(t)$ along the curve, let $X$ and $Y$ be parallel to their respective push-forward
\[
	X_{t}\parallel\varphi^{*}_{t}X_{0}\qquad Y_{t}\parallel\varphi^{*}_{t}Y_{0}.
\]
Then the angle between $X$ and $Y$ is constant along $\gamma$.
\end{Prop}

\begin{proof}
 As the curve $\gamma$ can be arbitrarily reparametrized and the conformal symmetries admit a group structure, it is sufficient to prove that the angle between $X$ and $Y$ is the same at $\gamma(0)$ and $\gamma(t)$ for an arbitrarily chosen $t\in I$. The angle between $X$ and $Y$ at $\gamma(0)$ is given by
\[
	\cos(\alpha(0))=\frac{g(X_{0},Y_{0})}{\sqrt{\left|g(X_{0},X_{0})g(Y_{0},Y_{0})\right|}}.
\]
Thus, we may easily calculate via (\ref{eqn:confsym}) the angle at $\gamma(t)$ with $f_{X}$ and $f_{Y}$ being the functions along $\gamma$ satisfying $X_{t}=f_{X}(t) \varphi^{*}_{t}X_{0}$ and $Y_{t}=f_{Y}(t)\varphi^{*}_{t}Y_{0}$.

\begin{eqnarray*}
	\cos(\alpha(t))
	 & = & \frac{g(X_{t},Y_{t})}{\sqrt{\left|g(X_{t},X_{t})g(Y_{t},Y_{t})\right|}}\\
	 & = & \frac{e^{t\Phi}f_{X}f_{Y}g(X_{0},Y_{0})}{e^{t\Phi}f_{X}f_{Y}
	\sqrt{\left|g(X_{0},X_{0})g(Y_{0},Y_{0})\right|}}=\cos(\alpha(0))
\end{eqnarray*}
\end{proof}

We will call such vector fields, which are related by the flow $\varphi_{t}$ and thus the angle between them is preserved, \textbf{conformally Lie transported} along the integral curves of $\xi$. For these we can calculate the Lie derivative along $\xi$ directly through its definition. First we get for the pullback of conformally Lie transported vector fields 
\[
	\varphi_{t*}X=f_{X}(t)X,
\]
again for an arbitrary smooth function $f_{X}$. Thus,
\[
	L_{\xi}X=\lim_{t\to 0}\frac{1}{t}(\varphi_{t*}X-X)=\lim_{t\to 0}\left(\frac{f_{X}(t)-1}{t}\right)X=
	\frac{df_{X}}{dt}X\parallel X.
\]
We will see that especially all lightlike vector fields and---even more important---their projections onto the hyperplanes perpendicular to $\xi$ are conformally Lie transported along the integral curves of $\xi$.

By using the derivation rules of the Lie derivative, one can see that any multiple by a constant $c$ of a CVF with conformal factor $\Phi$ is again a CVF with conformal factor $\Psi=c\Phi$:
\[
	L_{\eta}g=L_{c\xi}g=c\Phi g=\Psi g.
\]
This means that a CVF on $M$ induces an equivalence class of CVF's, the members of which are constant multiples of each other. This also corresponds to the fact that the set of CVF's on a manifold $M$ form a Lie algebra.

Now consider a world model with a CVF $\xi$ proportional to the observer field $V$, i.e.\ there is some function $f\colon M\to\mathbb{R}$ such that
\[
	\xi=f\cdot V,\qquad f>0.
\]
$f$ will be called \textbf{connecting function} in the following. World models with a globally defined CVF proportional to the observer field are also called \textbf{conformally stationary}. We will allow $f$ or $\xi$ to be defined only on a submanifold of $M$. We state the following proposition:

\begin{Prop}\label{prop:phiunddf}
On some open subset of a world model, let $\xi$ be a CVF that is parallel to the observer field $V$ with connecting function $f$. Then the conformal factor $\Phi$ is given by
\[
	\Phi=2df(V).
\]
\end{Prop}
\begin{proof} We insert twice the observer field into the conformal equation and make use of (\ref{eqn:liedec}).
\[
	(L_{\xi}g)(V,V)=\Phi g(V,V)=-\Phi
\]
\[
	\Rightarrow\ -\Phi=(L_{fV}g)(V,V)=f(L_{V}g)(V,V)+2df(V)g(V,V)
\]
Because of $0=L_{V}(g(V,V))=(L_{V}g)(V,V)+2g(\underbrace{L_{V}V}_{=0},V)=(L_{V}g)(V,V)$ we have
\[
	2df(V)=\Phi.
\]
\end{proof}
Obviously, the connecting function $f$ is only determined up to a constant factor for a given class of CVF's.

Since only the derivative of the connecting function enters the conformal equation, one could think that with $f$ giving rise to a CVF along $V$, also $f+c$ with $c=const\neq0$ would give rise to the same CVF. But one can easily check that this is not the case in general. One would have
\[
	\Phi g=L_{(f+c)V}g=\Phi g+cL_{V}g,
\] 
which is only true if $L_{V}g=0$, i.e.\ the observer field is a Killing field.

Now we prove that in the direction of the observer field there exists precisely one equivalence class of CVF's.

\begin{Prop}\label{prop:cvfunique}
Let $U\subseteq M$ and $W\subseteq U$ be open subsets of a world model $(M,g,V)$. Let $\xi=fV$ be a CVF defined on $U$, such that $L_{\xi}g=\Phi g$ and $\eta=h\xi=hfV$ another CVF defined on $W$, such that $L_{\eta}g=\Psi g$. Then $h$ is constant on $W$, that means $\xi$ and $\eta$ belong to same equivalence class of CVF's.
\end{Prop}

\begin{proof} Assume that $h$ is non-constant. Then by Prop. \ref{prop:phiunddf}, $\Phi=2df(V)$ and
\[
	\Psi=2d(hf)(V)=2fdh(V)+h\Phi
\]
holds. Thus by (\ref{eqn:liedec}) we have
\[
	L_{h\xi}g=hL_{\xi}g+dh\vee g(\xi,\cdot)=h\Phi g+dh\vee g(\xi,\cdot)=\Psi g.
\] 
This can only be true if
\[
	dh\vee g(V,\cdot)=2dh(V)g(\cdot,\cdot).
\]
This yields a contradiction as one immediately sees by introducing an orthonormal basis $\{V,E_{\alpha}\}$ on $W$, such that $E_{\alpha}\perp V$ ($\alpha=1,2,3$). Inserting $E_{\alpha}$ leads to
\[
	dh(E_{\alpha})g(V,\cdot)=2dh(V)g(E_{\alpha},\cdot).
\]
Since $\{V=E_{0},E_{\alpha}\}$ is an orthonormal basis this can only hold for $dh=0$. \end{proof}

The following corollary is obvious but has far-reaching consequences: we may use it to ``glue'' together CVF's that are just defined on open subsets to yield a unique global equivalence class of CVF's.

\begin{corollary}\label{cor:cvfglobal}
Let $\{U_{i}\}_{i\in J}$ be a locally-finite open covering of a world model $(M,g,V)$. Let $\{\xi_{i}\}_{i\in J}$ be a family of CVF's proportional to the observer field $V$, each defined on the open set $\xi_{i}\in TU_{i}\ \ \forall\,i\in J$. Then there is a unique equivalence class of CVF's $\xi\in TM$ proportional to the observer field defined globally on $(M,g,V)$. 
\end{corollary}   

\begin{proof} Obviously it follows from prop.\ \ref{prop:cvfunique}, that two vector fields $\xi_{i}$ and $\xi_{j}$ belong to the same equivalence class of CVF's on the intersection of the corresponding $U_{i}$ and $U_{j}$ for which $U_{i}\cap U_{j}\neq\emptyset$. Since this is valid for all $i,j\in J$, all the $\xi_{i}$ belong to the same equivalence class of CVF's, which we name $\xi$. As $M$ can be covered by the $U_{i}$'s, $\xi$ is defined globally on $M$. \end{proof}

In fact, this is valid for any covering of the world model, since any cover has a locally-finite refinement due to the paracompactness of the manifold $M$ (e.g.\ \cite{Conlon2001}, Lemma 1.4.8).

\subsection*{Light Signals}
As we will deal with the redshift of light rays modeled by null geodesics of a world model as well as with parallaxes of the view lines of an observer---which are also null geodesics---we have to establish some definitions on light signals, messages and null geodesics on a Lorentzian manifold. We mostly follow the definitions in \cite{Perlick1990}.

\begin{Def}\label{def:lichts}
\emph{(Set of light signals)} The set of light signals $\mathcal{C}$ on a world model $(M,g,V)$ are sections of future pointing lightlike geodesics in $[0,1]$-parametrisation:
\[
	\mathcal{C}:=\{\lambda:[0,1]\to M\ |\ \lambda \textrm{ is a } C^{\infty}\textrm{-map with tangential vector } 
	K:=\partial/\partial\lambda
\]
\[
	\textrm{ satisfying } \nabla_{K}K=0,\ g(K,K)=0,\ g(K,V)<0\} 
\]
\end{Def}

The endpoints of a light signal may be regarded as the sending event of one observer and the receiving event of another observer. However, for any two points on the manifold the existence of a light signal between them is not guaranteed because they are not necessarily timelikely separated.

\begin{Def}\label{def:jacobi}
\emph{(Jacobi field)} $J$ is called a Jacobi field along a lightlike geodesic with tangential vectorfield $K$ iff \begin{equation}\label{eqn:jac}
	\nabla_{K}\nabla_{K}J=R(K,J)K.
\end{equation}
\end{Def}
It is well-known that a Jacobi field $J$ provides an infinitesimal variation for the lightlike geodesic that is the integral curve of $K$. This means that the Jacobi field always points to an infinitesimally close geodesic. Thus, if the Jacobi field can be extended to points off the lightlike geodesic, it is the infinitesimal generator of a variation of the geodesic.

\begin{Def}\label{def:infmes}
\emph{(Infinitesimal message)} On a world model $(M,g,V)$, let $\lambda\in\mathcal{C}$ and $J$ be a vector field along $\lambda$. $J$ is called an infinitesimal message along $\lambda$ if the following four conditions hold:
\begin{enumerate}
	\item[(i)] $J$ is a Jacobi field,
	\item[(ii)] $g(K,\nabla_{K}J)=0$,
	\item[(iii)] $J_{\lambda(0)}=V_{\lambda(0)}$,
	\item[(iv)] $J_{\lambda(1)}\parallel V_{\lambda(1)}$.
\end{enumerate}
\end{Def}

In Definition \ref{def:infmes}, (i) expresses the fact that $J$ describes another geodesic (which is infinitesimal close to $\lambda$), (ii) ensures that this geodesic is lightlike, too, whereas (iii) and (iv) express the fact that it connects the same two observers as $\lambda$ does. Additionally, it is common to choose $J_{\lambda(0)}$ to be equal to $V_{\lambda(0)}$ in (iii). One could also choose it to be parallel, but with this convention one can show that along each light signal there is a unique infinitesimal message and the factor connecting $J_{\lambda(1)}$ and $V_{\lambda(1)}$ relates to the redshift along the signal; both provided that the light signal is \textbf{regular} \cite{Perlick1990}. A light signal is called regular if it is free of conjugate points. And by definition (e.g.\ \cite{O'Neill1983}, chap.\ 10) there exists a pair of conjugate points $p$ and $q$ along a geodesic $\sigma$ iff there is a nonzero Jacobi field along $\sigma$ that vanishes at $p$ and $q$. For light signals the following holds \cite{Perlick1990}: a light signal $\lambda\in\mathcal{C}$ is regular if and only if $\lambda(0)$ is not conjugate to $\lambda(1)$ along $\lambda$. In the following, we will only consider regular light signals for which a unique infinitesimal message exists along them. We do this for two reasons: first from a physical point of view we consider the ray-optical approximation to be valid, i.e.\ we speak about light rays and not waves. In doing so we imply that there is no focusing effect in the world model under consideration, which is known to correspond to conjugate points. Second from a mathematical point of view we consider rather light signals, than inextendible lightlike geodesics, i.e.\ $\mathcal{C}$ only contains sections of lightlike geodesics that satisfy the regularity condition above. This implies that we can cover any section of a lightlike geodesic containing a pair of conjugate points by two regular light signals: let $\sigma\colon[a,b]\to M$ be a lightlike geodesic and let $p,q\in [a,b]$ with $a<p<q<b$ be a pair of conjugate points along $\sigma$. Obviously the light signal $\lambda_{0}\in\mathcal{C}$ given by $\lambda_{0}(0)=p$ and $\lambda_{0}(1)=q$ is not regular -- but the light signals $\lambda_{1}\in\mathcal{C}$ and $\lambda_{2}\in\mathcal{C}$, given by $\lambda_{1}(0)=a$, $\lambda_{1}(1)=c$ and $\lambda_{2}(0)=c$, $\lambda_{2}(1)=b$ respectively, with $p<c<q$ but otherwise arbitrary, are regular, and $\lambda_{0}\subset\sigma\subset\lambda_{1}\cup\lambda_{2}$.   

\begin{Def}\label{def:mes}
\emph{(Message)} On a world model $(M,g,V)$, let $\gamma_{1}\colon I_{1}\to M$ and $\gamma_{2}\colon I_{2}\to M$ be two integral curves of $V$, with $I_{1}=[0,t]$ and $I_{2}$ being two real intervals. A $2$-surface $\mathcal{F}\subset M$ defined by the $C^{\infty}$-map
\[
	\mathcal{F}\colon [0,1]\times[0,t]\to M,\ (s,\tau)\mapsto\mathcal{F}(s,\tau)
\]
is called a message from $\gamma_{1}$ to $\gamma_{2}$ of temporal ($\gamma_{1}$)-duration $t$ iff
\begin{itemize}
	\item[(i)] $\mathcal{F}(\.,\tau)\in\mathcal{C}\ \ \forall\,\tau\in I_{1}$,
	\item[(ii)] For all $s\in[0,1]$, $\mathcal{F}(s,\.)$ is an integral curve of some Jacobi field $J(s,\.):=J_{\mathcal{F}(s,\.)}$, which is an infinitesimal message for any light signal considered in (i),
	\item[(iii)] $\mathcal{F}(0,\.)=\gamma_{1}|_{I_{1}}$ and $\mathcal{F}(1,\.)=\gamma_{2}|_{I_{2}}$.
\end{itemize}
\end{Def}

Here, the message maps $I_{1}$ diffeomorphically onto $I_{2}$.
For an arbitrary pair of observers $(\gamma_{1},\gamma_{2})$ neither the existence nor the uniqueness of a message between them is guaranteed. But as long as an infinitesimal message exists along a light signal, there will also exist a message, at least in an open neighborhood of the light signal to which the infinitesimal message can be extended.

Due to the definition above one can assert that a message is a lightlike geodesic variation, the endpoints of which vary along the integral curves of $V$. Thus a message consists of a family of lightlike geodesics all connecting the same two observers.

Any message $\mathcal{F}$ is a submanifold with boundary of $M$, which generally can be immersed as we allow for self-intersections. However, we will consider them being embedded from now on by splitting a message into shorter, non-intersecting messages. For any message $\mathcal{F}$, the Jacobi field $J$ and the lightlike tangential vector field $K$ to every light signal of the message form two pointwise linearly independent vector fields tangential to the message. Their Lie bracket can generally be calculated from (ii) in def.\ (\ref{def:infmes}),
\[
	0=g(K,\nabla_{K}J)=g(K,\nabla_{K}J)-\frac{1}{2}\nabla_{J}g(K,K)=g(K,[K,J])
\]
\[
	\Rightarrow [K,J]\parallel K.
\]
Hence $K$ and $J$ generate a subalgebra of $\mathcal{X}(M)$, which follows from Frobenius' theorem.

\section{The theorem of Hasse and Perlick}
The theorem of Hasse and Perlick states that the following four conditions are equivalent for a world model $(M,g,V)$ (see also tab. \ref{fig:hpt}).
\begin{itemize}
	\item[(i)] There is a CVF proportional to $V$ on $M$.
	\item[(ii)] The shear vanishes and the one-form $g(\nabla_{V}V,\.)-\frac{\Theta}{3}g(V,\.)$ is closed.
	\item[(iii)] $V$ admits a redshift potential.
	\item[(iv)] The world model is parallax-free.
\end{itemize}
(i)$\Leftrightarrow$(ii) is well-known and already Oliver and Davis gave a proof in \cite{Oliver1977}. However, we will fully work out the proof since to our best knowledge it has not yet been conducted in invariant notation. For (i)$\Leftrightarrow$(iii) we will give a simplified proof (compared to \cite{Hasse1988}) pointing out especially the connection of the CVF to the existence of a conserved quantity along lightlike geodesics using Noether's theorem. (i)$\Leftrightarrow$(iv) can also be simplified compared to the proof given in \cite{Hasse1988}; we will mimic the proof given in \cite{Perlick1990}, but conduct it in a straightfoward fashion using again Noether's theorem.

\begin{table}[htbp]
	\centering
		\begin{tabular}{ccccc}
			\fbox{Red-shift potential} & \huge{$\stackrel{\textrm{\scriptsize{\ref{thm:cf}}}}{\Longleftrightarrow}$} 
			& \fbox{$L_{\xi}g=\Phi g$} & 
			\huge{$\stackrel{\textrm{\scriptsize{\ref{prop:parakonf1}, \ref{prop:zurueck}}}}{\Longleftrightarrow}$}
			& \fbox{Parallax-freeness} \\
			 & & & & \\
			 & & \huge{$\Updownarrow$} \scriptsize{\ref{prop:shear}} & & \\
			 & & & & \\
			 & & \fbox{$\sigma=0$, $d\rho=0$} & & \\
			 & & & & 
		\end{tabular}
	\caption{Hasse--Perlick theorem}
	\label{fig:hpt}
\end{table}

\subsection*{Vanishing of shear}
The fact that a CVF proportional to the observer field leads to the vanishing of shear and vice versa is a well-known fact (e.g.\ \cite{Duggal1999} and \cite{Oliver1977}). It provides a connection between global and local properties of the world model and gives rise to the following proposition:

\begin{Prop}\label{prop:shear}
If $(M,g,V)$ is a world model, the following two properties are equivalent.
\begin{enumerate}
	\item[(i)] There is a vector field $\xi$ on $M$ such that $\xi=f\cdot V$ with a scalar function $f>0$ and $L_{\xi}g=\Phi g$. That means $\xi$ is a CVF proportional to $V$.
	\item[(ii)] $\sigma=0$ and the one-form $\rho:=g(\nabla_{V}V,\cdot)-\frac{\Theta}{3}g(V,\cdot)$ is
	closed (i.e.\ $d\rho=0$).
\end{enumerate}
\end{Prop} 

What attracts attention here is the fact that the existence of a CVF implies the vanishing of shear without any further necessary condition. The vanishing of the shear tensor alone does not imply the existence of a CVF parallel to the observer, but one has the additional condition that the exterior derivative of some one-form vanishes. This is because the CVF implies a conformal symmetry---which is basically a global condition---whereas the shear vanishes due to the shear tensor being zero at any point of the manifold -- which is an essentially pointwise condition. Thus, one has to find a way to ``integrate'' the vanishing of shear to a global condition. This leads to the problem of constructing a connecting function $f$ from the vanishing of shear alone, which is not possible. In demanding $d\rho=0$, one can use Poincar\'e's lemma to find a function $\tilde{f}$ defined on any convex subset of the manifold such that $\rho=d\tilde{f}$. We will see that $\tilde{f}=\ln f$ with the connecting function $f$. Since any Lorentzian manifold can be covered by a family of convex sets (\cite{O'Neill1983}, chap.\ 5), the local CVF's---each defined on a convex set---can be glued together to construct a global equivalence class of CVF's due to prop.\ (\ref{prop:cvfunique}) and its corollary.

\begin{proof} We present a proof following the idea in \cite{Oliver1977}, but using invariant notation and working it out fully.

(i)$\Rightarrow$(ii): We make use of the decomposition of $g(\cdot,\nabla_{\cdot}V)$ into the kinematical invariants (\ref{eqn:kininv}). We like to derive an equation relating $\Phi$, $f$ and $\Theta$. To this end, we first compute the trace of the conformal equation:
\begin{equation*}
	\textrm{tr}(L_{\xi}g)=\textrm{tr}(\Phi g).
\end{equation*}
Making use of (\ref{eqn:liedec}), (\ref{eqn:symkininv}), the definition of the expansion $\Theta$ and $\textrm{tr}(g)=2$, we obtain
\[
	\textrm{tr}(L_{\xi}g)=\textrm{tr}(fL_{V}g+df\vee g(V,\cdot))=2\Phi
\]
\[
	\Rightarrow\ \textrm{tr}(2f\cdot\textrm{sym}(g(\cdot,\nabla_{\cdot}V))+df\vee g(V,\cdot))=2\Phi
\]
\begin{equation}\label{eqn:phi1}
	\Rightarrow\ 2f\Theta+2df(V)=2\Phi.
\end{equation}
From proposition \ref{prop:phiunddf} we get
\begin{equation}\label{eqn:phi2}
	\Phi=2df(V).
\end{equation}
Combining (\ref{eqn:phi1}) and (\ref{eqn:phi2}) yields
\begin{equation}\label{eqn:phi3}
	\Phi=\frac{2}{3}f\Theta.
\end{equation}
Now we can rewrite the conformal equation with the help of (\ref{eqn:phi3}) and using (\ref{eqn:liedec})
\[
	L_{\xi}g=L_{fV}g=fL_{V}g+df\vee g(V,\cdot)=\frac{2}{3}f\Theta g.
\]
Again, by (\ref{eqn:symkininv}) one obtains
\[
	2f\left(\frac{\Theta}{3}h+\sigma-\frac{1}{2}g(\nabla_{V}V,\cdot)\vee g(V,\cdot)\right)+df\vee g(V,\cdot)=
	\frac{2}{3}f\Theta g.
\]
By using the definition of the projection operator (\ref{eqn:proj}) and solving for $\sigma$ one obtains the following expression for the shear
\begin{equation}\label{eqn:shear2}
	\sigma(\cdot,\cdot)=\frac{1}{2}\left(g(\nabla_{V}V,\cdot)-\frac{df}{f}\right)\vee g(V,\cdot)-
	\frac{\Theta}{3}g(V,\cdot)\otimes g(V,\cdot).
\end{equation}
Inserting $V$ into (\ref{eqn:shear2}) yields
\[
	g(\nabla_{V}V,\cdot)+\frac{df(V)}{f}g(V,\cdot)=\frac{df}{f}+\frac{2}{3}\Theta g(V,\cdot).
\]
By carefully examining this equation one sees that $g(\nabla_{V}V,\cdot)$ is orthogonal to $g(V,\cdot)$, whereas $\frac{df(V)}{f}g(V,\cdot)$ and $\frac{2}{3}\Theta g(V,\cdot)$ are parallel to $g(V,\cdot)$. By projecting this equation onto the hyperplanes perpendicular to $g(V,\.)$, only $g(\nabla_{V}V,\cdot)$ and the component of $\frac{df}{f}$ orthogonal to $g(V,\.)$ remains, the latter of which is obviously $\frac{df}{f}+\frac{df(V)}{f}g(V,\cdot)$. Thus, one has
\[
	g(\nabla_{V}V,\cdot)=\frac{df}{f}+\frac{df(V)}{f}g(V,\cdot).
\]
Now, by (\ref{eqn:phi2}) and (\ref{eqn:phi3}) the following holds:
\begin{equation}\label{eqn:rhoundf1}
	\frac{df}{f}=g(\nabla_{V}V,\cdot)-\frac{\Theta}{3}g(V,\cdot)=\rho.
\end{equation}
Thus,
\begin{equation}\label{eqn:rhoundf2}
	\rho=d(\ln f)\ \Rightarrow\ d\rho=dd(\ln f)=0.
\end{equation}
Now by starting from (\ref{eqn:shear2}) and using (\ref{eqn:rhoundf1}) we can easily show that $\sigma$ vanishes
\[
	\sigma=\frac{1}{2}\left(g(\nabla_{V}V,\cdot)-\rho\right)\vee g(V,\cdot)-\frac{\Theta}{3}g(V,\cdot)\otimes g(V,\cdot)
\]
\[
	\Rightarrow\ \sigma=\frac{1}{2}\frac{\Theta}{3}g(V,\cdot)\vee g(V,\cdot)-\frac{\Theta}{3}g(V,\cdot)\otimes 
	g(V,\cdot)=0.
\]
(ii)$\Rightarrow$(i): Let $\sigma=0$ and $d\rho=d\left(g(\nabla_{V}V,\cdot)-\frac{\Theta}{3}g(V,\cdot)\right)=0$. By Poincar�'s lemma there is a scalar function $\ln f$ such that $d(\ln f)=\rho$ ($f>0$). We choose 
\[
	\Phi:=\frac{2}{3}f\Theta\qquad\textrm{and}\qquad\xi:=fV.
\]
Now we calculate $L_{\xi}g$, using (\ref{eqn:proj}), (\ref{eqn:symkininv}) and the definitions above:
\[
	L_{\xi}g=fL_{V}g+df\vee g(V,\cdot)
\]
\[
	=2f\left(\frac{\Theta}{3}h-\frac{1}{2}g(\nabla_{V}V,\cdot)\vee g(V,\cdot)\right)+df\vee g(V,\cdot)
\]
\[
	= \Phi g+\underbrace{\frac{2}{3}f\Theta g(V,\cdot)\otimes g(V,\cdot)-fg(\nabla_{V}V,\cdot)\vee g(V,\cdot)+
	df\vee g(V,\cdot)}_{=:A}.
\]
By the definition of $\rho$ one can now show that $A$ vanishes
\[
	A=2f(\frac{\Theta}{3}g(V,\cdot)\otimes g(V,\cdot)-\frac{1}{2}g(\nabla_{V}V,\cdot)\vee g(V,\cdot)+
	\frac{1}{2}\frac{df}{f}\vee g(V,\cdot)).
\]
\[
	=2f(\frac{\Theta}{3}g(V,\cdot)\otimes g(V,\cdot)-\frac{1}{2}g(\nabla_{V}V,\cdot)\vee g(V,\cdot)+
	\frac{1}{2}(g(\nabla_{V}V,\cdot)-\frac{\Theta}{3}g(V,\cdot))\vee g(V,\cdot))=0
\]
Hence we recover the conformal equation
\[
	L_{\xi}g=\Phi g.
\]
\end{proof}

The following corollary of this proposition was proven by Oliver and Davis, too (see \cite{Oliver1977}). We present it here in invariant notation and work out the proof. It gives the first hint that if a CVF exists, the rotation cannot be totally arbitrary, but has to satisfy some integrability conditions. The equations one gets are similar to Raychaudhuri's equations for $\omega$, which describe the evolution of $\omega$ along the observer field in terms of the covariant derivative (e.g.\ \cite{Kar2007}). Since $\omega$ and $g(\nabla_{V}V,\cdot)$ are differential forms with $g(\nabla_{V}V,V)=\omega(V,\.)=0$, due to eqn.(\ref{eqn:lieform}) their Lie derivative relates also to the ``$V$ component'' of their exterior derivative.

\begin{corollary}\label{cor:intbed}
On a world model $(M,g,V)$ admitting a CVF $\xi$ proportional to the observer field $V$, the following two integrability conditions for the acceleration and the rotation must hold:
\begin{itemize}
	\item[(i)] $L_{V}\omega=\frac{\Theta}{6}\omega$
	\item[(ii)] $L_{V}g(\nabla_{V}V,\cdot)=\frac{1}{3}h(d\Theta-\frac{\Theta}{2}g(\nabla_{V}V,\cdot))$
\end{itemize}
Here, $h(\cdot)$ denotes the projection on the hyperplanes perpendicular to $V$.
\end{corollary}

\begin{proof} First we show that
\begin{equation}\label{eqn:vinnendv}
	V\rfloor dg(V,\cdot)=-\frac{1}{2}g(\nabla_{V}V,\cdot).
\end{equation}
Using a basis $\{E^{i}\}$ one obtains $V^{i}\nabla_{[j}V_{i]}=\frac{1}{2}(V^{i}\nabla_{j}V_{i}-V^{i}\nabla_{i}V_{j})$, which leads to $V\rfloor dg(V,\cdot)=\frac{1}{2}(g(\nabla_{\cdot}V,V)-g(\nabla_{V}V,\cdot))$ in invariant notaion. By (\ref{eqn:kininv}) we obtain $g(\nabla_{\cdot}V,V)=0$, which yields (\ref{eqn:vinnendv}).
 
For $\rho=g(\nabla_{V}V,\cdot)-\frac{\Theta}{3}g(V,\cdot)$ we get
\begin{equation}\label{eqn:drho}
	d\rho=dg(\nabla_{V}V,\cdot)-\frac{d\Theta}{3}\wedge g(V,\cdot)-\frac{\Theta}{3}dg(V,\cdot)=0.
\end{equation}
By ``wedging'' this equation with $g(V,\cdot)$ one obtains
\begin{equation}\label{eqn:drho2}
	dg(\nabla_{V}V,\cdot)\wedge g(V,\cdot)=\frac{\Theta}{3}dg(V,\cdot)\wedge g(V,\cdot).
\end{equation}
(i) We calculate $L_{V}\omega$. From the definition of the rotation we get the general expression
\[
	\omega=-dg(V,\.)-\frac{1}{2}g(V,\.)\wedge g(\nabla_{V}V,\.),
\]
which implies
\[
	d\omega=\frac{1}{2}dg(\nabla_{V}V,\.)\wedge g(V,\.)-\frac{1}{2}g(\nabla_{V}V,\.)\wedge dg(V,\.).
\]
Inserting (\ref{eqn:drho2}) it follows that
\[
	d\omega=\frac{\Theta}{6}dg(V,\cdot)\wedge g(V,\cdot)-\frac{1}{2}g(\nabla_{V}V,\.)\wedge dg(V,\.).
\]
By (\ref{eqn:lieform}) we obtain
\[
	L_{V}\omega=V\rfloor d\omega+d(V\rfloor\omega),
\]
and due to $V\rfloor\omega=\omega(V,\.)=0$ this yields
\[
	L_{V}\omega=V\rfloor(\frac{\Theta}{6}dg(V,\cdot)\wedge g(V,\cdot)-\frac{1}{2}g(\nabla_{V}V,\.)\wedge dg(V,\.))
\]
\[
	=\frac{\Theta}{6}\left[\left(V\rfloor dg(V,\.)\right)\wedge g(V,\.)+dg(V,\.)\wedge
	\left(V\rfloor g(V,\.)\right)\right]-
\]
\[
	-\frac{1}{2}\left[\left(V\rfloor g(\nabla_{V}V,\.)\right)\wedge dg(V,\.)-g(\nabla_{V}V,\.)
	\wedge\left(V\rfloor dg(V,\.)\right)\right].
\]
Due to $V\rfloor g(V,\.)=g(V,V)=-1$, $V\rfloor g(\nabla_{V}V,\.)=g(\nabla_{V}V,V)=0$ and (\ref{eqn:vinnendv}) it follows that
\[
	L_{V}\omega=\frac{\Theta}{6}\left[-\frac{1}{2}g(\nabla_{V}V,\.)\wedge g(V,\.)-dg(V,\.)\right]=\frac{\Theta}{6}\omega.
\]
(ii) We calculate $L_{V}g(\nabla_{V}V,\.)$ using again (\ref{eqn:lieform}). This yields
\[
	L_{V}g(\nabla_{V}V,\.)=V\rfloor dg(\nabla_{V}V\.)+d(\underbrace{V\rfloor g(\nabla_{V}V,\.)}_{=0}).
\]
Inserting (\ref{eqn:drho}) and (\ref{eqn:vinnendv}) leads to
\[
	L_{V}g(\nabla_{V}V,\.)=(V\rfloor\frac{d\Theta}{3})\wedge g(V,\.)-\frac{d\Theta}{3}(V\rfloor g(V,\.))+\frac{\Theta}{3}
	V\rfloor dg(V,\.)
\]
\[
	=\frac{1}{3}(d\Theta(V)g(V,\.)+d\Theta)+\frac{\Theta}{6}g(\nabla_{V}V,\.)=\frac{1}{3}h(d\Theta-
	\frac{\Theta}{2}g(\nabla_{V}V,\.)).
\]
\end{proof}

This corollary gives rise to the following theorem. It was already proven by Oliver and Davis \cite{Oliver1977} and also by Perlick \cite{Perlick1990}. The fundamental statement of this theorem is that world models with a CVF proportional to the observer field cannot both rotate and expand if the acceleration vanishes.

\begin{Thm}\label{thm:exporrot}
If there is a CVF $\xi$ on a world model $(M,g,V)$ with $\xi\parallel V$ and zero acceleration ($\nabla_{V}V=0$), then the expansion vanishes
\[
	\Theta=0
\]
or the rotation vanishes
\[
	\omega=0.
\] 
\end{Thm} 

\begin{proof} From (\ref{eqn:drho2}) in corollary \ref{cor:intbed} we find for vanishing acceleration
\begin{equation}\label{eqn:drho3}
	\frac{\Theta}{3}g(V,\.)\wedge dg(V,\.)=0.
\end{equation}
Computing the wedge product of the definition of $\omega$ with $g(V,\.)$ we find 
\begin{equation}\label{eqn:womega}
	g(V,\.)\wedge dg(V,\.)=-g(V,\.)\wedge\omega.
\end{equation}
Inserting (\ref{eqn:womega}) into (\ref{eqn:drho3}) and taking the inner product of the equation with the observer field $V$ yields
\[
	\frac{\Theta}{3}\omega=0.
\] 
Thus, $\Theta$ or $\omega$ must vanish.
\end{proof}

This theorem lays important constraints upon the construction of cosmological models with rotation and expansion. If we like to construct a world model with vanishing shear ($\sigma=0$), $\Theta\neq0$ and $\omega\neq0$, we can either set the acceleration to zero ($\nabla_{V}V=0$) or demand that there is a CVF proportional to the observer field ($\exists\,\xi\parallel V$). In the first case we loose the properties of parallax-freeness and the existence of a redshift potential, but the world lines of the matter are geodesics. In the second case the matter does not move along geodesics but we have a conformally stationary spacetime. 

\subsection*{Red Shift} We like to recover the properties of the redshift of a conformally stationary world model via a conserved quantity. So we first state Noether's theorem in a general and more informal version.
\begin{Thm}\label{thm:noether}
\emph{(Noether's Theorem I)} For every continuous symmetry of the action there is exactly one quantity that is conserved along the physical trajectories.
\end{Thm}
Null geodesics like the trajectories of light rays can be obtaind from a variational principle as has been shown in \cite{Perlick2000} for a Hamiltonian treatment or in \cite{O'Neill1983}, chap.\ 10, where the action functional and the allowed variational vector fields are given expicitly. One can easily write down a Lagrangian in terms of a local chart that leads to geodesic curves in the manifold under consideration. Let
\[
	s\colon \mathbb{R}\to M\qquad s\mapsto x(s)
\]
be a curve in the manifold $M$ and $x^{a}(s)$ its components with respect to a local chart. Thus the tangential vector field to the curve is given by
\[
	\dot{x}(s)=\frac{dx(s)}{ds}=\dot{x}^{a}(s)\frac{\partial}{\partial x^{a}}.
\]
We define a Lagrangian by
\[
	\mathcal{L}\colon TM\to\mathbb{R}\qquad\mathcal{L}(x,\dot{x})=\frac{1}{2}g_{ab}(x)\dot{x}^{a}\dot{x}^{b}.
\]
If this Lagrangian is constrained to the nullcone, i.e.
\[
	\mathcal{L}=0,
\]
the tangential vectors of the curve are lightlike and we obtain lightlike geodesics for the physical trajectories. The nullcone forms a seven-dimensional submanifold $N\subset TM$ (see \cite{Sachs1977}, chap.\ 5.6). Thus, one gets all lightlike geodesics of the manifold (in terms of a local chart) from introducing Lagrangian multipliers in the Euler--Lagrange equations. This is also known as the d'Alembert--Lagrange principle (e.g.\ \cite{Arnold1993}, chap.\ 2.5) 
\begin{equation}\label{eulerlagrange}
	\frac{d}{ds}\left(\frac{\partial\mathcal{L}}{\partial\dot{x}^{a}}\right)-
	\frac{\partial\mathcal{L}}{\partial x^{a}}=\mu(s)\frac{\partial\mathcal{L}}{\partial\dot{x}^{a}}.
\end{equation}
First we can calculate the generalized momentum $k=k_{c}dx^{c}$ for $\dot{x}^{c}$
\[
	k_{c}=\frac{\partial\mathcal{L}}{\partial\dot{x}^{c}}=g_{ab}(x)\frac{\partial\dot{x}^{a}}{\partial\dot{x}^{c}}
	\dot{x}^{b}=g_{ab}(x)\delta^{a}_{c}\dot{x}^{b}=g_{cb}\dot{x}^{b}.
\]
From this we can also obtain a Hamiltonian $H(x,k)$ by applying the Legendre transform:
\[
	H:\,T^{*}M\to\mathbb{R}\qquad H(x,k)=k_{a}\dot{x}^{a}(x,k)-\mathcal{L}(x,\dot{x}(x,k)).
\]
To calculate this quantity, we have to solve the velocities for the momenta
\[
	k_{a}=g_{ab}\dot{x}^{b}\ \Rightarrow\ \dot{x}^{a}=g^{ab}k_{b}.
\]
Hence
\[
	H(x,k)=k_{a}k_{b}g^{ab}-\frac{1}{2}g_{ab}g^{ac}g^{bd}k_{c}k_{d}=\frac{1}{2}g^{ab}(x)k_{a}k_{b}.
\]
To evaluate the Euler--Lagrange equations we express the operator $\frac{d}{ds}$ in terms of the covariant derivative and the components of the Levi--Civita connection (Christoffel symbols)
\[
	\frac{d}{ds}(k_{c})=\dot{x}^{a}\partial_{a}(k_{c})=\dot{x}^{a}\nabla_{a}(k_{c})+
	\dot{x}^{a}\Gamma^{b}_{ac}k_{b},
\]
where $\dot{x}^{a}\nabla_{a}=\nabla_{\dot{x}}=\frac{D}{ds}$.

It is well-known that the partial derivative of the metric tensor can also be expressed in terms of Christoffel symbols
\[
	\frac{\partial}{\partial x^{c}}(g_{ab})=\Gamma^{d}_{cb}g_{da}+\Gamma^{d}_{ca}g_{db}.
\] 
Thus (\ref{eulerlagrange}) becomes
\[
	g_{ab}\nabla_{\dot{x}}\dot{x}^{b}+g_{db}\Gamma^{d}_{ac}\dot{x}^{b}\dot{x}^{c}-\frac{1}{2}\dot{x}^{b}\dot{x}^{c}
	(\Gamma^{d}_{ab}g_{dc}+\Gamma^{d}_{ac}g_{db})=\mu(s)g_{ab}\dot{x}^{b}
\]
\[
	\Rightarrow\ g_{ab}\nabla_{\dot{x}}\dot{x}^{b}+\dot{x}^{b}\dot{x}^{c}\Gamma^{d}_{a[c}g_{b]d}=\mu(s)g_{ab}\dot{x}^{b}.
\]
The second term in this equation vanishes due to symmetries of the indices. Hence we recover the pregeodesic equation for the velocities
\[
	\nabla_{\dot{x}}\dot{x}=\mu(s)\dot{x},
\]
which leads to lightlike geodesic equations by affine reparametrisation
\[
	\nabla_{K}K=0
\]
for
\[
	\dot{x}^{a}\frac{\partial}{\partial x^{a}}=K\qquad\textrm{and}\qquad\ k_{a}dx^{a}=g(K,\.).
\]
If we now consider conformal transformations
\[
	g(x)\to g^{\prime}(x^{\prime})=e^{\Phi(x)}g(x),
\]
we see that they preserve the nullcone and the Lagrangian becomes
\[
	\mathcal{L}\to\mathcal{L}^{\prime}=\frac{1}{2}e^{\Phi(x)}g_{ab}(x)\dot{x}^{a}\dot{x}^{b}.
\]
This primed Lagrangian leads to the same lightlike pregeodesics, up to reparametrisation, as the unprimed one. We can assume the conformal transformation to be non-isometric (i.e.\ $\Phi\neq 0$) as it is well-known that isometries are symmetries for all geodesics. One easily calculates the conjugate momentum for $\mathcal{L}^{\prime}$ to
\[
	k_{a}=\Phi g_{ab}\dot{x}^{b}.
\]
Therefore we have to evaluate the Euler--Lagrange equations
\[
	\frac{d}{ds}\left(\frac{\partial\mathcal{L}}{\partial\dot{x}^{a}}\right)-
	\frac{\partial\mathcal{L}}{\partial x^{a}}=\mu(s)\Phi g_{ab}\dot{x}^{b},
\]
which---by a similar calculation as the one above---results in
\[
	\nabla_{\dot{x}}\dot{x}=\frac{\nabla\Phi}{\Phi}\mathcal{L}-\frac{g(\nabla\Phi,\dot{x})}{\Phi}\dot{x}
	+\mu(s)\dot{x},
\]
where we denote the gradient of $\Phi$ by $\nabla\Phi$ such that $d\Phi=g(\nabla\Phi,\.)$.

Introduction of the nullcone condition $\mathcal{L}=0$ yields
\[
	\nabla_{\dot{x}}\dot{x}=\left(\mu-\frac{g(\nabla\Phi,\dot{x})}{\Phi}\right)\dot{x}.
\] 
This is basically again the lightlike geodesic equation by means of an appropriate affine transformation. What we emphasized here in terms of a Lagrangian treatment is the well-known fact that conformal transformations map lightlike pregeodesics to lightlike pregeodesics (see e.g.\ \cite{Beem1996}). Therefore a one-parameter group of conformal symmetries on a manifold $M$ leaves the nullcone submanifold $N\subset TM$ invariant, and according to Noether's theorem there must be a conserved quantity along the lightlike geodesics corresponding to this symmetry. This conserved quantity can be computed from the infinitesimal version of the symmetry, i.e.\ from the CVF inducing it. So we can now state Noether's theorem in a more explicit version (see \cite{Arnold1993}, chap.\ 3, theorems\ 1 and\ 3). 

\begin{Thm}\label{thm:noetherII}
\emph{(Noether's Theorem II)} A Lagrangian system $(M,L)$ admits a one-parameter group of symmetries $\Phi_{t}$ if and only if $I=\langle\xi,V\rangle$ is a first integral of the equations of motion.
\end{Thm} 
By a Lagrangian system $(M,L)$ we mean a $C^{\infty}$-manifold $M$ and a Lagrangian $L\colon TM\to\mathbb{R}$ from which the equations of motion can be derived. $\langle\.,\.\rangle$ is an inner product on $TM$, in our case $\langle\.,\.\rangle=g(\.,\.)$. $V$ consists of the tangential vector fields along the physical trajectories and $\xi|_{p}=\frac{d}{dt}\Phi_{t}(p)|_{t=0}$ is the vector field generating the symmetry, in our case the CVF. As a first integral of the equations of motion we consider a quantity that is constant along the physical trajectories. This will be used now for the case of conformal transformations, CVF's and light signals.

\begin{Thm}\label{thm:cf}
\emph{(Conservation of conformal frequency)} On a world model $(M,g,V)$ a vector field $\xi$ proportional to the observer field is a CVF iff the quantity $g(\xi,K)$ is conserved along any light signal $\lambda\in\mathcal{C}$. $g(\xi,K)$ is called conformal frequency (CF) in the following. This means, that the following two conditions are equivalent: 
\begin{itemize}
	\item[(i)] $L_{\xi}g=\Phi g$,
	\item[(ii)] $\nabla_{K}(g(\xi,K))=0\ \textrm{or}\ g(\xi,K)=\textrm{\emph{const} along any } \lambda$.
\end{itemize}
\end{Thm}

\begin{proof} (i)$\Rightarrow$(ii): For a given CVF this is shown easily by direct calculation using only the lightlike geodesic equation $\nabla_{K}K=0$, the fact that $K$ is a null vector and the decomposition $\nabla_{K}\xi=\nabla_{\xi}K+L_{\xi}K$:
\[
	\nabla_{K}(g(\xi,K))=g(\nabla_{K}\xi,K)+g(\xi,\nabla_{K}K)=g(\nabla_{\xi}K,K)+g(L_{\xi}K,K)
\]
\[
	=\frac{1}{2}\nabla_{\xi}(g(K,K))+\frac{1}{2}L_{\xi}(g(K,K))-\frac{1}{2}(L_{\xi}g)(K,K)
\]
\[
	=-\frac{1}{2}(L_{\xi}g)(K,K)=-\frac{\Phi}{2}g(K,K)=0\qquad
\]
(ii)$\Rightarrow$(i): This is difficult to show by direct calculation but is obvious from Noether's theorem (\ref{thm:noetherII}). \end{proof}

Nevertheless, (ii)$\Rightarrow$(i) can be proved by direct calculation using the Newman--Penrose (NP) formalism for CVF's introduced by Ludwig et al., see \cite{Ludwig2001} and the references therein. From the conservation of the CF it follows that
\begin{equation}\label{view}
  g(\nabla_{K}\xi,K)=0,
\end{equation}
which does not contain any derivatives of the null vector field $K$ anymore. Thus, this equation must be valid for any lightlike vector $K|_{p}$ at an arbitrary point $p\in M$. In a sufficiently small neighborhood of $p$ this vector can be extended to a null vector field, for which $\nabla_{\xi}(g(K,K))=0$ is inevitably true. This results in
\[
	0=\frac{1}{2}\nabla_{\xi}(g(K,K))-g(\nabla_{K}\xi,K)=g(\nabla_{\xi}K-\nabla_{K}\xi,K)=g(L_{\xi}K,K)
\] 
for every lightlike vector field $K$, and in particular for the null-tetrad vectors in the NP formalism. The conditions $g(L_{\xi}K_{(i)},K_{(i)})=0$ must hold for every tetrad vector $K_{(i)}$ with $i\in\{0,1,2,3\}$ under spin-boost transformations and null rotations (null-tetrad gauge). These transformations for the tetrad $\{K_{(i)}\}=\{l,k,m,\bar{m}\}$ (with $g(l,k)=-1$, $g(m,\bar{m})=1$ and all other products zero) are
\begin{equation}\label{spin-boost}
	l\to A\.l\ \ \ \ \ \ k\to A^{-1}\.k\ \ \ \ \ \ m\to e^{i\Theta}m\ \ \ \ \ \ \bar{m}\to e^{-i\Theta}\bar{m}\ ,
\end{equation}
with the real functions $A$ and $\Theta$ on $M$ and
\begin{equation}\label{nullrot}
	l\to l\ \ \ \ \ \ m\to m+\bar{c}l\ \ \ \ \ \ \bar{m}\to\bar{m}+cl\ \ \ \ \ \ k\to k+cm+\bar{c}\bar{m}+c\bar{c}l
\end{equation}
with the complex function $c$.

For a given vector field $\xi=fV$ parallel to the observer field $V$ one can choose the spin gauge (\ref{spin-boost}) such that $V=\frac{1}{\sqrt{2}}(l+k)$ and the general formula $L_{\xi}V=-df(V)V$ yields the conformal factor $\Phi:=\frac{1}{2}df(V)$. By further demanding the invariance of (\ref{view}) under null rotations (\ref{nullrot}) one obtains for the tetrad vectors
\[
	L_{\xi}l=-\frac{\Phi}{2}l\qquad L_{\xi}k=-\frac{\Phi}{2}k\qquad L_{\xi}m=-\frac{\Phi}{2}m\qquad 
	L_{\xi}\bar{m}=-\frac{\Phi}{2}\bar{m}.
\]
It can be easily checked that this leads to the conformal equation $L_{\xi}g=\Phi g$ for the metric.

The following remark is in order: as we check for the definitions \ref{def:infmes} and \ref{def:mes} of an (infinitesimal) message we see that also the Jacobi field $J$ defined on any message obeys the condition $g(\nabla_{K}J,K)=0$ for the lightlike vector field $K$ of the light signals the message consists of. Thus, the Jacobi field $J$ is a CVF on a message, regarded as a $2$-surface $\mathcal{F}$, i.e.\ a CVF for the metric induced on $\mathcal{F}$ by the embedding in $M$ (inner metric of $\mathcal{F}$). This is obvious since the flow of $J$ maps the lightlike geodesics in $\mathcal{F}$ into each other. In the sense of the Lagrangian treatment above the flow of $J$ on any message is a symmetry for a Lagrangian constrained to the null-cone $N_{\mathcal{F}}\subset N\subset TM$ of $\mathcal{F}$. The fact that such a symmetry always exists on a message corresponds to the fact that a message $\mathcal{F}$ can be regarded as a $2$-dimensional spacetime itself. As it is well-known that $2$-dimensional Lorentzian manifolds are conformally flat, there always exists a CVF on them.

We will now relate the theorem of CF conservation to the notion of a redshift potential found in \cite{Hasse1988}. Since the CVF is proportional to the observer field with connecting function $f$, it follows from theorem \ref{thm:cf} that the quantity $f\.g(V,K)$ is conserved along lightlike geodesics. It is well-known that $\nu_{p}=|g(V_{p},K_{p})|$ is the (physical) frequency of a light ray with tangential vector field $K$ that crosses an integral curve of the observer field at a point $p\in M$, measured by the observer at $p$. Thus, the conservation of the CF can be considered as the reason for the change of the physical frequency along a null geodesic -- also known as the redshift.

The redshift along a light signal is usually defined as the ratio of the frequency change to the emitted frequency. For a light signal connecting two points $p,q\in M$ the redshift is thus given by

\begin{equation}\label{eqn:redshift}
	z_{p\to q}=\frac{\Delta \nu}{\nu_{p}}=\frac{\nu_{p}-\nu_{q}}{\nu_{p}},
\end{equation} 
which is negative if the received frequency is higher than the emitted frequency (blue shift). A notion that is more suitable for our purposes is the redshift function of a world model \cite{Perlick1990}.

\begin{Def}\label{def:redshiftfunc}
\emph{(Red-shift function)} A function $r\colon\mathcal{C}\to\mathbb{R}$ defined by
\[
	\exp{r(\lambda)}:=\frac{g(K(0),V_{\lambda(0)})}{g(K(1),V_{\lambda(1)})}
\]
is called a redshift function on $(M,g,V)$.
\end{Def}
A redshift function relates to $z_{p\to q}$ via $z_{\lambda(0)\to\lambda(1)}=\exp{r(\lambda)}-1$. Let $s$ be the affine parameter of any light signal $\lambda$ (running from 0 to 1), then one can easily check \cite{Perlick1990} that the redshift function obeys
\[
	r(\lambda)=\ln g(V,K)|_{\lambda(0)}-\ln g(V,K)|_{\lambda(1)}=-\int^{1}_{0}
	\frac{g(\nabla_{K}V,K)}{g(V,K)}|_{\lambda(s)}ds.
\]
We are now ready to define the following:

\begin{Def}\label{def:redshiftpot}
\emph{(Red-shift potential)} A redshift function $r$ on a world model $(M,g,V)$ is said to have a potential $f\colon M\to\mathbb{R}$---called redshift potential---if the redshift between two points in $M$ only depends on the value of $f$ at these points, i.e.\
\[
	r(\lambda)=\ln f(\lambda(0))-\ln f(\lambda(1))
\] 
holds for every light signal $\lambda$.
\end{Def}
A redshift potential provides another means of describing conformally stationary world models:

\begin{Prop}\label{prop:redshiftpot}
On a world model $(M,g,V)$, there is a CVF $\xi=fV$ proportional to the observer field $V$ if and only if the connecting function $f$ is a redshift potential. 
\end{Prop}

\begin{proof} We show that the existence of a redshift potential is equivalent to the conservation of conformal frequency:
\[
	0=\nabla_{K}(g(\xi,K))=\frac{d}{ds}(f\.g(V,K))=g(V,K)\frac{df}{ds}+f\frac{d}{ds}(g(V,K))
\]
\[
	\Leftrightarrow\ \frac{d(\ln f)}{ds}=\frac{\frac{d}{ds}(g(V,K))}{g(V,K)}=\frac{g(\nabla_{K}V,K)}{g(V,K)}
\]
\[
	\Leftrightarrow\ -\int_{\lambda}d(\ln f)=-\int^{1}_{0}\frac{g(\nabla_{K}V,K)}{g(V,K)}|_{\lambda(s)}ds=r(\lambda)
\]
\[
	r(\lambda)=\ln f(\lambda(0))-\ln f(\lambda(1))\ \ 
\]
\end{proof}
On a conformally stationary world model, with the 1-form $\rho=d(\ln f)$ we can write for the redshift function:
\begin{equation}\label{eqn:redshiftCVF}
r(\lambda)=\int_{\lambda}\rho.
\end{equation}

Thus, we conclude in this case by def.\ (\ref{eqn:redshift}) and (\ref{eqn:redshiftCVF}) that
\[
	z_{p\to q}=\frac{f(q)}{f(p)}-1.
\]
Hence the redshift stays the same if we multiply $f$ by a constant, i.e.\ the redshift only depends on the equivalence class of CVF's.

We also see that the redshift is isotropic iff $z_{p\to q}$ depends only on the spatial distance of $p$ and $q$. A sufficient condition for this is the vanishing of the acceleration $\nabla_{V}V=0$. In this case due to (\ref{eqn:redshiftCVF}) we have 
\[
	r(\lambda)=-\int_{\lambda}\frac{\Theta}{3}g(V,\.).
\] 
However, zero acceleration is not necessary for isotropy in redshift, as claimed in \cite{Perlick1990}. One can construct conformally stationary world models with $\nabla_{V}V\neq 0$ which exhibit an isotropic redshift; but the acceleration is not completely arbitrary: $\int_{\lambda}\rho$ has to be a function of the spatial distance of the points $p$ and $q$, connected by $\lambda$ only.

\subsection*{Parallaxes}
First we define three notions of parallax-freeness following \cite{Perlick1990}.

\begin{Def}\label{def:spf}
\emph{(Parallax-free in the strong sense)} $(M,g,V)$ is called parallax-free in the strong sense (sPf) iff for any three observers $\gamma_{0},\gamma_{1},\gamma_{2}$ the following holds: the angle between $\gamma_{1}$ and $\gamma_{2}$ as seen on the celestial sphere of $\gamma_{0}$ is constant over time (proper time of $\gamma_{0}$). 
\end{Def}

\begin{Def}\label{def:wpf}
\emph{(Parallax-free in the weak sense)} $(M,g,V)$ is called parallax-free in the weak sense (wPf) iff for any three observers $\gamma_{0},\gamma_{1},\gamma_{2}$ the following holds true: if $\gamma_{0}$ at one instant of time (proper time of $\gamma_{0}$) sees $\gamma_{1}$ and $\gamma_{2}$ in the same spatial direction so he sees them in the same spatial direction at every instant of time. 
\end{Def}

\begin{Def}\label{def:mpf}
\emph{(Parallax-free in the mathematical sense)} $(M,g,V)$ is called parallax-free in the mathematical sense (mPf) iff for any message $\mathcal{F}$ in the world model, the observer field restricted to $\mathcal{F}$ is tangential to $\mathcal{F}$, i.e.\ on the message the observer field $V$ is a linear combination of the Jacobi field $J$ and the lightlike vector field $K$. Hence we have
\begin{equation}\label{eqn:parfree1}
	J(\tau,s)-vV_{\mathcal{F}(\tau,s)}\parallel K(\tau,s)\qquad\forall (\tau,s)\in[0,t]\times[0,1]
\end{equation}
with an appropriate function $v=v(\tau,s)$ on any message.
\end{Def}

Obviously sPf implies wPf. In fact it is also obvious that wPf implies mPf: one observer $\gamma_{0}$ seeing two other observers $\gamma_{1}$ and $\gamma_{2}$ in the same spatial direction---at one instant in time---means they are ``behind'' each other in the sense that the light signal $\lambda_{0}$ starting at $\gamma_{2}$ and ending at $\gamma_{0}$ crosses the world line of $\gamma_{1}$ for some parameter $s\in[0,1]$.
 If we assume that this light signal is given by $\tau=0$ in a message between $\gamma_{2}$ and $\gamma_{0}$ (i.e.\ $\lambda_{0}=\mathcal{F}(0,\.)$), the world line $\gamma_{1}$ has to be contained in $\mathcal{F}$ for all $\tau\in[0,t]$ in order for $\gamma_{2}$ to be ``behind'' $\gamma_{1}$ at all later times. Since this must be true for any three observers, this gives eqn. (\ref{eqn:parfree1}).
Not so obvious is the fact that the converse statement also holds. We will prove it along with the fact that parallax-freeness is equivalent to the existence of a CVF. Hereby we use the proof structure of table \ref{fig:beweispara}. 

\begin{table}[htbp]
	\centering
		\begin{tabular}{lcr}
			\fbox{sPf} & \huge{$\stackrel{\textrm{\scriptsize{trivial}}}{\Longrightarrow}$} & \fbox{wPf} \\
			 & & \\
			\huge{$\Uparrow$} \scriptsize{Prop. \ref{prop:zurueck}} & & \scriptsize{trivial} \huge{$\Downarrow$}  \\
			 & & \\
			\fbox{$L_{\xi}g=\Phi g$} & \huge{$\stackrel{\textrm{\scriptsize{Prop. \ref{prop:parakonf2}}}}
			{\Longleftarrow}$} & \fbox{mPf}\\
			 & & 
		\end{tabular}
	\caption{Proof structure of parallax-freeness}
	\label{fig:beweispara}
\end{table}

\begin{Prop}\label{prop:parakonf1}
If a world model $(M,g,V)$ is parallax-free in the mathematical sense, then on any message there is a CVF proportional to the observer field.
\end{Prop}

\begin{proof} In the following we will suppress the coordinates $s$ and $\tau$ for simplicity. First we orthogonally project both sides of (\ref{eqn:parfree1}) onto $K$, which yields
\[
	g(K,J)-vg(K,V)=0\ \Rightarrow\ v=\frac{g(K,J)}{g(K,V)}.
\] 
Since we only consider future pointing lightlike geodesics, $g(K,V)$ is negative. For $g(K,J)$ we conclude from (ii) in def.\ \ref{def:infmes} that $g(K,J)|_{s}$ is constant along any light signal $\lambda$ and therefore  $g(K,J)=g(K(0),V_{\lambda(0)}<0$. Hence we have $v>0$ on the whole message $\mathcal{F}$.

Now projecting (\ref{eqn:parfree1}) with $g(K,\nabla_{K}\ \.)$ yields, since $\lambda$ is a null geodesic,
\[
	g(K,\nabla_{K}J)-g(K,\nabla_{K}(vV))\parallel g(K,\nabla_{K}K)=0
\]
\[
	\Rightarrow\ g(K,\nabla_{K}J)=g(K,\nabla_{K}(vV))=\nabla_{K}(g(K,vV)).
\]
Again by def.\ \ref{def:infmes} we have $g(K,\nabla_{K}J)=0$. Hence
\begin{equation}\label{eqn:parfree3}
	\nabla_{K}(g(K,vV))=0
\end{equation}
holds. By theorem \ref{thm:cf}, $\xi:=vV$ then is a CVF proportional to the observer field on any message $\mathcal{F}$. \end{proof}

We have constructed CVF's proportional to the observer field essentially on the whole world model, but each of them only defined on a message. What is left to prove is that this leads in fact to a global CVF. This is provided by the following proposition.

\begin{Prop}\label{prop:parakonf2}
If a world model $(M,g,V)$ is parallax-free in the mathematical sense, then there is a unique equivalence class of CVF's proportional to the observer field on $M$.
\end{Prop}

\begin{proof} By parallax-freeness we have a CVF proportional to $V$ on any message. As $V$ is tangential to any message and since $[K,J]\parallel K$, one also has $[K,V]\parallel K$ due to eqn.\ (\ref{eqn:parfree1}). Thus, the observer field and the lightlike vector field $K$ of any message generate a closed subalgebra of $\mathcal{X}(M)$. Using Frobenius' theorem, there is a codimension two foliation of any open subset $U_{i}\subset M$, the plaques of which are messages. We choose the subsets $U_{i}$ such that they form a locally-finite cover of $M$. Then due to prop.\ \ref{prop:cvfunique} the CVF on any $U_{i}$ is unique. That means in this case, we get the same CVF if we choose another foliation, with another connecting function $v$. Now one can glue together the CVF on any $U_{i}$ by means of the corollary to prop.\ \ref{prop:cvfunique}. \end{proof}

The next proposition provides the way back, from a CVF proportional to the observer field to the condition that the world model is sPf. In proving this we show that all three definitions of parallax-freeness are in fact equivalent as well as equivalent to conformal stationarity.

\begin{Prop}\label{prop:zurueck}
If a world model $(M,g,V)$ provides a CVF $\xi\parallel V$ then $(M,g,V)$ is parallax-free in the strong sense.
\end{Prop} 
\begin{proof} Choose an arbitrary observer $\gamma_{0}$ receiving two arbitrary messages $\mathcal{F}_{1}$ and $\mathcal{F}_{2}$ of arbitrary duration $t$. Let the observers $\gamma_{1}$ and $\gamma_{2}$ be the sources of these messages, respectively. Then the null vectors $K_{1}(1,\tau)$ and $K_{2}(1,\tau)$ belong to the messages and provide two lightlike vector fields along $\gamma_{0}$. The directions on the celestial sphere of $\gamma_{0}$ from which she receives the messages are given by the projection of $K_{1}$ and $K_{2}$ to the hyperplanes perpendicular to $V$. Hence
\[
	Y_{i}=K_{i}+g(K_{i},V)V\qquad i=1,2
\] 
are two spacelike vector fields along $\gamma_{0}$. Since the vector fields $K_{i}$ each belong to a unique message, they are conformally Lie transported along $\gamma_{0}$, i.e.\ $L_{\xi}K_{i}=\alpha_{i}K_{i}$ for some functions $\alpha_{i}$. This is because $[K,J]\parallel K$ on any message and $J\parallel V$ on $\gamma_{0}$. The projections $Y_{i}$ are conformally Lie transported, too -- as a straightforward calculation shows, using $L_{\xi}g=2df(V)g$ and $L_{\xi}V=-df(V)V$
\[
	L_{\xi}Y_{i}=\alpha_{i}K_{i}+L_{\xi}(g(K_{i},V)V)=\alpha_{i}K_{i}-df(V)g(K_{i},V)V+
\]
\[
	+2df(V)g(K_{i},V)V+\alpha_{i}g(K_{i},V)V-df(V)g(K_{i},V)V=\alpha_{i}Y_{i}.
\]
Thus, prop.\ \ref{prop:angle} tells us that the angle between $Y_{1}$ and $Y_{2}$ remains constant. As this is valid for any observer in $(M,g,V)$, the world model is parallax-free in the strong sense. \end{proof}

\section{Causality}
As has been shown in \cite{Wald1984}, the existence of a global time function is a necessary and sufficient condition for a spacetime to be stably causal. Since a CVF proportional to the observer field naturally provides a function $f$ on $M$---the connecting function---this raises the question under which circumstances it represents a global time function.

\begin{Thm}\label{thm:stablecausal}
\emph{(Stable causality)} A time orientable spacetime is stably causal iff there is a differentiable function $f$ globally on $M$, such that its gradient is timelike and past pointing everywhere.
\end{Thm}

Since we deal with world models only, time orientability is always given because of the existence of an observer field. One can now easily deduce two propositions which give some requirements to the connecting function $f$ and the kinematical invariants. Although these requirements are sufficient for the world model under consideration being stably causal, they are in no way necessary. The following proposition \ref{prop:cvfgrad} can be regarded as a special case of proposition \ref{prop:cvfcausal}, defining a class of conformally stationary spacetimes that is comprised of those world models for which the gradient of the connecting function is always parallel to the observer field. One then sees that in this class of world models the expansion must not change its sign. It is also well-known (see \cite{Sachs1977}) that for such world models (which are synchronizable, i.e.\ $g(V,\.)\propto df$) the rotation vanishes.

\begin{Prop}\label{prop:cvfgrad}
Let $(M,g,V)$ be a world model which admits a proper CVF $\xi$ proportional to the observer field $V$. Let $f$ be the connecting function, i.e.\ $\xi=fV$. If the gradient of the connecting function is proportional to $V$ and nowhere zero, then the spacetime $(M,g)$ is stably causal.
\end{Prop}

\begin{proof} We denote by $\nabla f$ the gradient of the connecting function $f$, such that $df=g(\nabla f,\.)$. As $f>0$ by definition, $\nabla(-f)$ is timelike and past pointing:
\[
	g(\nabla(-f),\nabla(-f))\propto g(-V,-V)=-1<0\qquad\Rightarrow\textrm{timelike}
\]
\[
	g(\nabla(-f),V)\propto-g(V,V)=1>0\qquad\Rightarrow\textrm{past pointing}
\]
Thus, $(M,g)$ is stably causal by theorem \ref{thm:stablecausal}. \end{proof}

In case the gradient of $f$ is not proportional to $V$, the spacetime may be stably causal if the gradient remains timelike and the expansion is nowhere zero ($|\Theta|>0$). However, this puts a constraint on the acceleration $\nabla_{V}V$.

\begin{Prop}\label{prop:cvfcausal}
Let $(M,g,V)$ be a world model which admits a proper CVF $\xi$ proportional to the observer field $V$. Let $f$ be the connecting function such that $\xi=fV$ and let $|\Theta|>0$. If for the acceleration and expansion
\[
	g(\nabla_{V}V,\nabla_{V}V)<\frac{\Theta^{2}}{9}
\]
holds, the spacetime $(M,g)$ is stably causal.
\end{Prop}

\begin{proof} Without loss of generality we may assume $f>1$ by choosing the right CVF from the equivalence class. This leads to $\ln f>0$. Hence we have
\[
	\rho=d(\ln f)=g(\nabla(\ln f),\.).
\]
This implies
\[
	g(\nabla(\ln f),V)=\rho(V)=g(\nabla_{V}V,V)-\frac{\Theta}{3}g(V,V)=\frac{\Theta}{3}.
\]
Thus we can choose $\ln f$ as a time function if $\Theta>0$ and $-\ln f$ if $\Theta<0$, such that its gradient is always past pointing.

The squared norm of $\nabla(\ln f)$ reads
\[
	g(\nabla(\ln f),\nabla(\ln f))=g(\nabla_{V}V,\nabla_{V}V)-\frac{\Theta^{2}}{9}.
\]
The spacetime $(M,g)$ is stably causal if this norm is strictly negative.
\end{proof}

\bibliography{literatur_alex}
\bibliographystyle{english_custom}
\end{document}